\documentclass[conference]{IEEEtran}
\IEEEoverridecommandlockouts
\usepackage{cite}
\usepackage{tikz}
\usepackage{soul}
\usepackage{amsmath,amssymb,amsfonts}
\usepackage{algorithmic}
\usepackage{graphicx}
\usepackage{hyperref}
\usepackage[table]{xcolor}
\usepackage{booktabs}
\usepackage{adjustbox}
\usepackage{titlesec}
\usepackage{textcomp}
\usepackage{xcolor}
\usepackage{adjustbox}
\usepackage{multirow}
\usepackage{array}
\usepackage{subcaption}
\usepackage{float}

\usepackage{url}
\usepackage{booktabs} 
\usepackage{caption} 
\usepackage{subcaption} 
\usepackage{array} 
\usepackage{kotex}



\def\BibTeX{{\rm B\kern-.05em{\sc i\kern-.025em b}\kern-.08em
    T\kern-.1667em\lower.7ex\hbox{E}\kern-.125emX}}
\begin{document}

\title{LatentTune: Efficient Tuning of High Dimensional Database Parameters via Latent Representation Learning*\\
}

\author{
\IEEEauthorblockN{Sein Kwon\textsuperscript{*}}
\IEEEauthorblockA{Computer Science\\
\textit{Yonsei University}\\
Seoul, Korea\\
{\footnotesize\url{seinkwon97@yonsei.ac.kr}}}
\and
\IEEEauthorblockN{Youngwan Jo}
\IEEEauthorblockA{Computer Science\\
\textit{Yonsei University}\\
Seoul, Korea\\
{\footnotesize\url{jyy1551@yonsei.ac.kr}}}
\and
\IEEEauthorblockN{Seungyeon Choi}
\IEEEauthorblockA{Computer Science\\
\textit{Yonsei University}\\
Seoul, Korea\\
{\footnotesize\url{tmddus1553@yonsei.ac.kr}}}
\and
\IEEEauthorblockN{Jieun Lee}
\IEEEauthorblockA{Computer Science\\
\textit{Yonsei University}\\
Seoul, Korea\\
{\footnotesize\url{jieun199624@yonsei.ac.kr}}}
\and
\IEEEauthorblockN{Huijun Jin}
\IEEEauthorblockA{Computer Science\\
\textit{Yonsei University}\\
Seoul, Korea\\
{\footnotesize\url{jinjuijun@yonsei.ac.kr}}}
\and
\IEEEauthorblockN{Sanghyun Park\textsuperscript{$\dagger$}}
\IEEEauthorblockA{Computer Science\\
\textit{Yonsei University}\\
Seoul, Korea\\
{\footnotesize\url{sanghyun@yonsei.ac.kr}}}
\thanks{\textsuperscript{*}First author. \textsuperscript{$\dagger$}Corresponding author.}
}

\maketitle

\begin{abstract}
As data volumes continue to grow, optimizing database performance has become increasingly critical, making the implementation of effective tuning methods essential. Among various approaches, database parameter tuning has proven to be a highly effective means of enhancing performance. Recent studies have shown that machine learning techniques can successfully optimize database parameters, leading to significant performance improvements.
However, existing methods still face several limitations. First, they require substantial time to generate large training datasets. Second, to cope with the challenges of high-dimensional optimization, they typically optimize only a subset of parameters rather than the full configuration space. Third, they often rely on information from similar workloads instead of directly leveraging information from the target workload. To address these limitations, we propose LatentTune, a novel approach that differs fundamentally from traditional methods. To reduce the time required for data generation, LatentTune incorporates a data augmentation strategy.
Furthermore, it constructs a latent space that compresses information from all database parameters, enabling the optimization of the full configuration space. In addition, LatentTune integrates external metric information into the latent space, allowing for precise tuning tailored to the actual target workload. Experimental results demonstrate that LatentTune outperforms baseline models across four workloads on MySQL and RocksDB, achieving up to 1332\% improvement for RocksDB and 11.82\% throughput gain with 46.01\% latency reduction for MySQL.
\end{abstract}

\begin{IEEEkeywords}
Database parameter tuning, Bayesian Optimization, Database Management Systems, Machine Learning
\end{IEEEkeywords}

\section{Introduction}
Recently, the volume of data has grown significantly, and the necessity for high-performance database management systems (DBMS) to process large amounts of data efficiently has increased \cite{DB_survey}. One of the approaches to improving DBMS performance is database parameter tuning \cite{DB_review}, which refers to the process of improving the performance of the database by adjusting the values of the parameters. Conventional database parameter tuning requires Database Administrators (DBAs) to modify parameter values; however it is difficult to tune parameters efficiently due to the complex relationship between numerous parameters. Furthermore, different databases have distinct parameters, and DBAs must account for these variations along with the diversity of workloads. To overcome these limitations, recent research has been conducted on automatic database parameter tuning using machine learning (ML) techniques, achieving improved performance in various workloads and DBMSs \cite{iTuned,Tuneful,OtterTune,ResTune,CDBTune,Qtune}. These studies have employed various ML-based optimization algorithms, including Bayesian Optimization (BO) \cite{BO} and reinforcement learning (RL) \cite{RL} to guide the tuning process more effectively. 

\noindent\textbf{Limitations of Existing Methods.} Machine learning models require large amounts of training data consisting of configurations and their corresponding performance metrics for effective training \cite{machinelearning}. In conventional approaches, such datasets are typically generated by sampling configuration values and evaluating their performance using benchmarking tools \cite{oltpbenchmark,sysbench,AuctionMark}.

\begin{table*}[!]
\caption{Top-10 parameters using SHAP and LASSO for RocksDB R90W10 and R50W50 workloads. Overlapped parameters are highlighted in bold.}
\label{tab:top-10-rocksdb}
\setlength{\tabcolsep}{2pt} 
\renewcommand{\arraystretch}{1.05}
\begin{tabular}{clcl}
    \hline
    \multicolumn{4}{c}{\textbf{R90W10}} \\ \hline
    \multicolumn{2}{c}{\textbf{SHAP (top-10)}}                     & \multicolumn{2}{c}{\textbf{Lasso (top-10)}}                    \\ \hline
    \multicolumn{2}{c}{\textbf{target\_file\_size\_base}}          & \multicolumn{2}{c}{compression\_type}                          \\
    \multicolumn{2}{c}{write\_buffer\_size}                        & \multicolumn{2}{c}{target\_file\_size\_multiplier}             \\
    \multicolumn{2}{c}{level0\_file\_num\_compaction\_trigger}     & \multicolumn{2}{c}{\textbf{num\_levels}}                       \\
    \multicolumn{2}{c}{min\_write\_buffer\_number\_to\_merge}      & \multicolumn{2}{c}{\textbf{target\_file\_size\_base}}          \\
    \multicolumn{2}{c}{compression\_ratio}                         & \multicolumn{2}{c}{bloom\_locality}                            \\
    \multicolumn{2}{c}{block\_size}                                & \multicolumn{2}{c}{\textbf{max\_background\_compactions}}               \\
    \multicolumn{2}{c}{\textbf{max\_background\_compactions}}      & \multicolumn{2}{c}{max\_background\_flushes}         \\
    \multicolumn{2}{c}{\textbf{num\_levels}}                       & \multicolumn{2}{c}{level0\_stop\_writes\_trigger}              \\
    \multicolumn{2}{c}{\textbf{memtable\_bloom\_size\_ratio}}      & \multicolumn{2}{c}{\textbf{memtable\_bloom\_size\_ratio}}      \\
    \multicolumn{2}{c}{\textbf{level0\_slowdown\_writes\_trigger}} & \multicolumn{2}{c}{\textbf{level0\_slowdown\_writes\_trigger}} \\ \hline
\end{tabular}
    \begin{tabular}{clcl}
    \hline
    \multicolumn{4}{c}{\textbf{R50W50}} \\ \hline
    \multicolumn{2}{c}{\textbf{SHAP (top-10)}}                        & \multicolumn{2}{c}{\textbf{Lasso (top-10)}}                 \\ \hline
    \multicolumn{2}{c}{\textbf{compression\_ratio}}                   & \multicolumn{2}{c}{\textbf{compression\_ratio}}             \\
    \multicolumn{2}{c}{write\_buffer\_size}                           & \multicolumn{2}{c}{\textbf{max\_backgound\_flushes}}        \\
    \multicolumn{2}{c}{max\_write\_buffer\_number}                    & \multicolumn{2}{c}{num\_levels}                             \\
    \multicolumn{2}{c}{min\_write\_buffer\_buffer\_number\_to\_merge} & \multicolumn{2}{c}{\textbf{targer\_file\_size\_multiplier}} \\
    \multicolumn{2}{c}{lwbel0\_file\_num\_compaction\_trigger}        & \multicolumn{2}{c}{bloom\_locality}                         \\
    \multicolumn{2}{c}{max\_bytes\_for\_level\_base}                  & \multicolumn{2}{c}{\textbf{memtable\_bloom\_size\_ratio}}   \\
    \multicolumn{2}{c}{compaction\_pri}                               & \multicolumn{2}{c}{level0\_stop\_writes\_trigger}           \\
    \multicolumn{2}{c}{\textbf{target\_file\_size\_multiplier}}       & \multicolumn{2}{c}{max\_background\_compactions}            \\
    \multicolumn{2}{c}{\textbf{level0\_stop\_writes\_trigger}}        & \multicolumn{2}{c}{level0\_slowdown\_writes\_trigger}       \\
    \multicolumn{2}{c}{\textbf{max\_background\_flushes}}             & \multicolumn{2}{c}{block\_size}                             \\ \hline
\end{tabular}
\vspace{-10pt}
\end{table*}

\noindent However, this process is time-consuming, making it difficult to collect data efficiently. For example, when we limited the benchmarking duration to 300 seconds per configuration in MySQL \cite{mysql}, generating a dataset of 1,000 configurations required substantial amounts of time, highlighting practical constraints in creating sufficiently large training datasets. 


Another challenge arises from the parameter selection process used to address high-dimensional optimization. 
To reduce the search space, existing BO-based methods~\cite{iTuned,Tuneful,OtterTune,ResTune} commonly adopt selection algorithms such as SHapley Additive exPlanations (SHAP)~\cite{SHAP} and LASSO~\cite{LASSO} to identify and optimize only the top-$k$ parameters that most significantly affect performance. 
RL-based approaches~\cite{Qtune,CDBTune}, on the other hand, utilize reinforcement learning algorithms such as Deep Deterministic Policy Gradient (DDPG)~\cite{DDPG} to cope with high-dimensional continuous spaces by learning effective representations of actions and states.

\begin{figure}[t]  
    \begin{minipage}{\columnwidth}  
        \centering
        \includegraphics[width=\columnwidth]{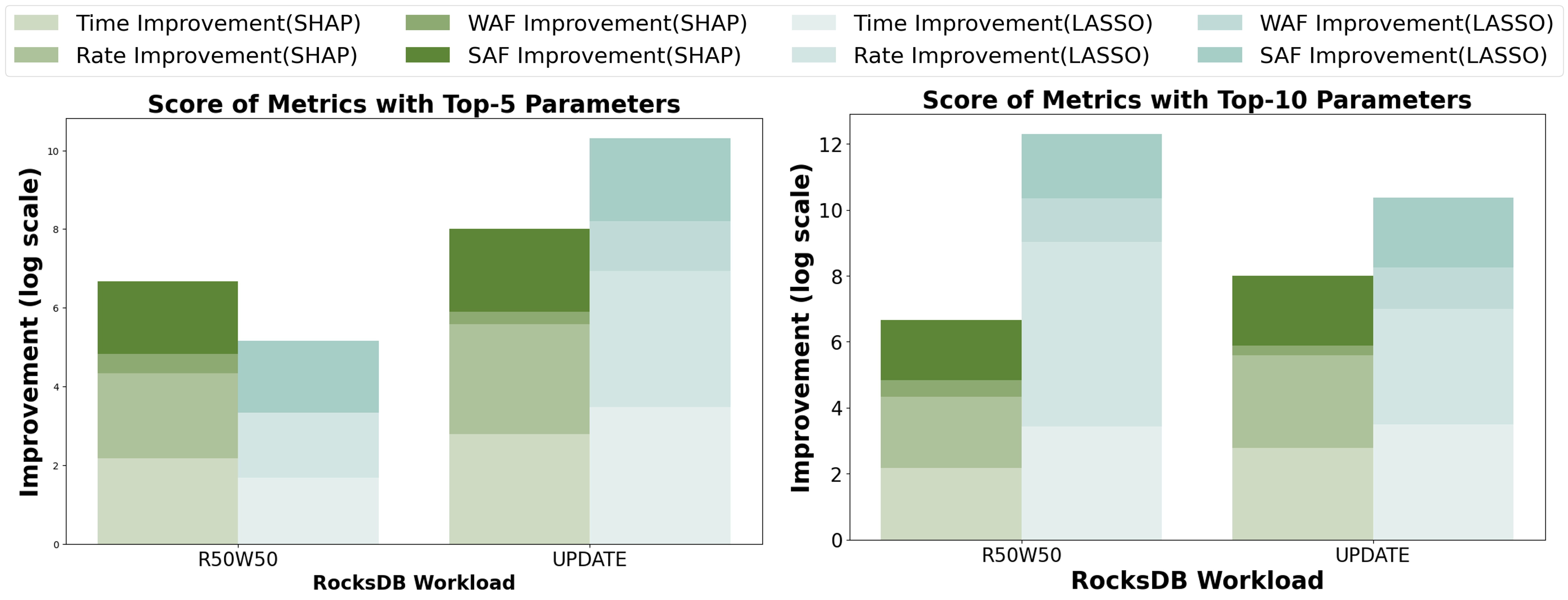}  
        \caption{Performance improvement with Top-5 and Top-10 Parameters on RocksDB.}
        \label{fig:top-param-rocksdb-performance}
    \end{minipage}
    \vspace{-20pt}
\end{figure}

\noindent However, the selected parameters can vary depending on the algorithm, even under the same workload. As shown in Table \ref{tab:top-10-rocksdb}, only three to five parameters overlap between the SHAP and LASSO algorithms for RocksDB~\cite{rocksdb}, and the ranking order also differs. Figure~\ref{fig:top-param-rocksdb-performance} presents the performance impact of tuning with RGPE~\cite{Facilitating}, one of our baseline models, using the top 5 and top 10 parameters. The results indicate that variations in parameter selection can lead to substantial differences in performance. Moreover, focusing only on the top-\textit{k} parameters may fail to capture critical interactions among the remaining parameters and overlook subtle but influential effects on database performance, ultimately reducing the stability of the tuning results. 

In addition to these challenges, adapting to diverse workloads remains a significant difficulty in database parameter tuning. Some existing methods attempt to identify the most similar workload from a data repository by computing its similarity to the target workload. However, this approach does not directly optimize for the target workload itself, and inaccurate workload mapping may lead to suboptimal tuning results due to the mismatch between the assumed and actual workload characteristics.
Even when two workloads appear similar based on internal metrics such as query latency or cache hit ratio, deeper structural differences ,including variations in query patterns, transaction sizes, and read/write ratios, can substantially affect which parameters are most critical for optimization. Since many internal metrics primarily capture low-level system behaviors rather than high-level workload structures, relying solely on past workload data may lead to ineffective tuning, ultimately degrading database performance. This highlights the need for a more target specific optimization approach that directly leverages performance feedback from the actual workload.

\noindent\textbf{Our Approach.} To address these limitations, we propose \textbf{LatentTune}, a framework that efficiently tunes high-dimensional database parameters through latent representation learning. When the training dataset is limited, LatentTune employs data augmentation techniques using Latin Hypercube Sampling (LHS) \cite{LHS} and TabNet \cite{TabNet}. Unlike conventional methods, our approach avoids the need for benchmarking to construct configuration and metric data pairs. To handle the challenge of high-dimensional parameter spaces, LatentTune leverages an AutoEncoder (AE)~\cite{AE} to compress all parameter configurations into a low-dimensional latent space, thereby enabling the joint optimization of all parameters. Furthermore, by injecting direct target-workload information into the latent space, LatentTune supports workload-specific tuning, removing the need to rely on similarity-based workload mapping. This integration leads to more accurate and effective tuning across diverse workload scenarios. Our key contributions are as follows:
\begin{itemize}
  \item LatentTune reduces the time required for dataset generation by employing data augmentation techniques.
  
  \item We introduce a novel framework, LatentTune, which addresses the high-dimensionality challenge in database parameter tuning by projecting the full set of configuration parameters into a compact latent space. This enables full-parameter optimization, in contrast to conventional methods that rely on selecting only a subset of parameters.
  
  \item LatentTune incorporates target workload information directly into the latent space, enabling workload-specific tuning that adapts to diverse workload characteristics.
  
  \item The experimental results demonstrate that LatentTune achieves superior performance through comparison with various baseline models.
\end{itemize}


\section{PRELIMINARIES}

\subsection{Automatic Database Parameter Tuning}

Database parameter tuning is a method of optimizing database parameters (such as memory allocation, disk storage, cache size etc.) to enhance database performance. The configuration of the database refers to the settings and parameters that determine the operation of the database system. If the parameters are specified as \( P \), the configuration can be defined as  \( \text{conf} = \{ P_1, \ldots, P_j \} \), where \textit{j} is the number of parameters. Figure \ref{fig:tuning process} depicts the basic pipeline of ML-based database parameter tuning. The overall flow of the tuning process can be described as follows:

\begin{figure}[t]  
    \begin{minipage}{\columnwidth}  
        \centering
        \includegraphics[width=\columnwidth]{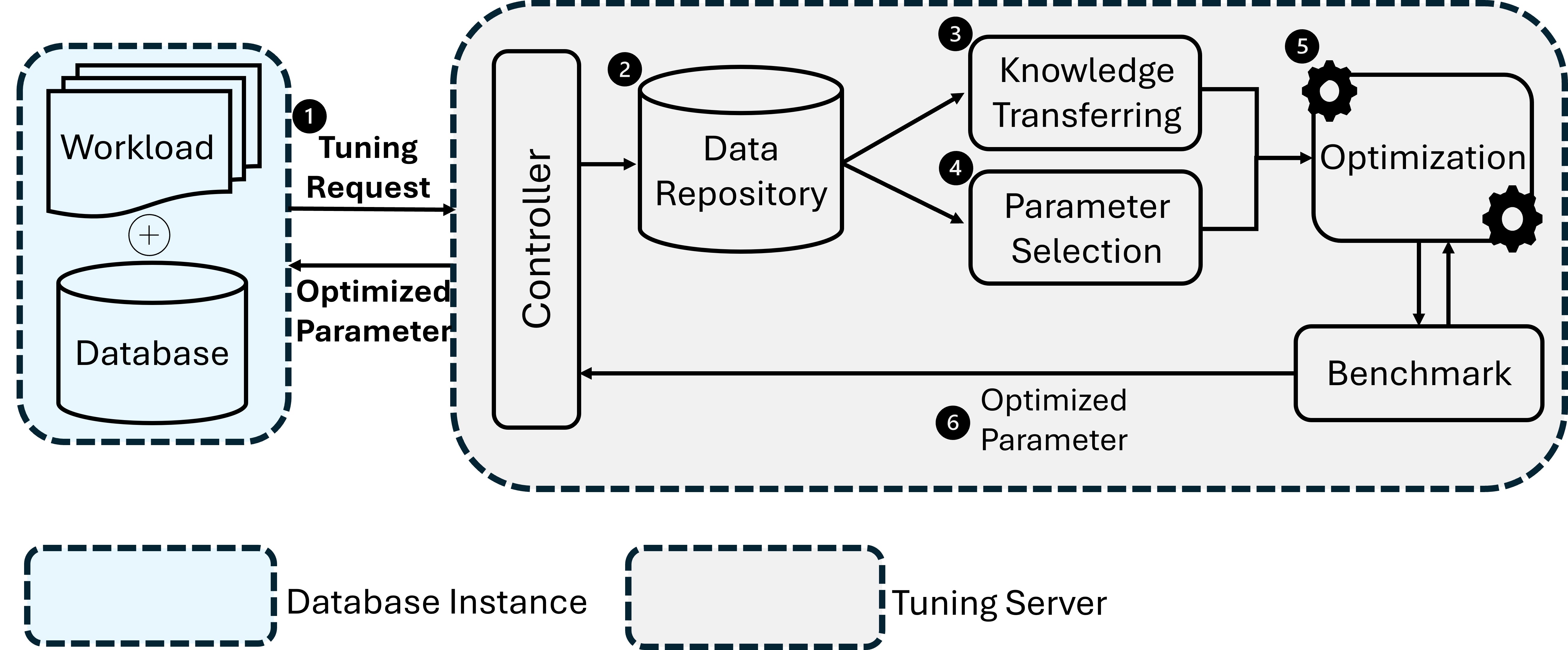}  
        \caption{Architecture of Database Parameter Tuning}
        \label{fig:tuning process}
    \end{minipage}
    \vspace{-15pt}
\end{figure}

\noindent{\textbf{\tikz[baseline=(char.base)] \node[shape=circle,draw,inner sep=0.5pt] (char) {\textbf{1}}; Tuning Request}.When the controller receives information about the DBMS and the workload to be tuned, it initiates the tuning process. \textbf{\tikz[baseline=(char.base)] \node[shape=circle,draw,inner sep=0.5pt] (char) {\textbf{2}}; Data Repository}. During the tuning process, the DBMS and workload information provided in the tuning request are stored in the data repository. \textbf{\tikz[baseline=(char.base)] \node[shape=circle,draw,inner sep=0.5pt] (char) {\textbf{3}}; Knowledge Transferring}. To optimize the target workload, the process calculates the similarity between the target and stored workloads in the data repository using a distance metric such as Euclidean distance \cite{ED}. The most similar workload information is then utilized to guide the tuning process. The objective of this procedure is to facilitate efficient tuning across diverse workloads. \textbf{\tikz[baseline=(char.base)] \node[shape=circle,draw,inner sep=0.5pt] (char) {\textbf{4}}; Parameter Selection}. To address the challenge of optimization in high-dimensional search spaces, a parameter selection algorithm identifies the parameters that most strongly influence database performance. \textbf{\tikz[baseline=(char.base)] \node[shape=circle,draw,inner sep=0.5pt] (char) {\textbf{5}}; Optimization}. The optimization algorithm, such as BO or RL, optimizes the top-\textit{k} parameters that have a significant impact on database performance ({\tikz[baseline=(char.base)] \node[shape=circle,draw,inner sep=0.5pt] (char) {4};) and information about the target workload ({\tikz[baseline=(char.base)] \node[shape=circle,draw,inner sep=0.5pt] (char) {3};). \textbf{\tikz[baseline=(char.base)] \node[shape=circle,draw,inner sep=0.5pt] (char) {\textbf{6}}; Optimized Parameter}. The optimized parameters are passed to the controller, which then applies these parameters in the actual database.}

\subsection{Related Work}

Automatic database parameter tuning techniques are divided into two categories: search-based methods \cite{bestconfig} and ML-based methods \cite{OtterTune,ResTune,Qtune,CDBTune}. Search-based methods allow tuning without detailed information about the internals of the database system, and ML-based methods tune the parameters using ML models.

{\textbf{Search-based method.}}
Search-based tuning is a technique for optimizing database performance by finding database parameters according to predefined rules. BestConfig~\cite{bestconfig} uses Divide and Diverge Sampling (DDS) for parameter sampling in high-dimensional parameter spaces and Recursive Bound and Search (RBS) to find optimal configurations in the bounded space around it. However, search-based methods can become unstable with low-quality data and rely on heuristic techniques that are often time-consuming.

\titlespacing*{\subsubsection}{0pt}{*1.0}{*0.5}{\textbf{ML-based method.}}
ML-based methods use ML models to optimize parameters. For instance, OtterTune \cite{OtterTune} selects parameters that are important for performance to optimize in a high-dimensional search space and uses BO with a Gaussian Process (GP) \cite{GP} to enhance performance. However, since the set of parameters optimized by OtterTune varies depending on the parameter selection algorithm used, it may result in unstable performance.
CDBTune \cite{CDBTune} uses Reinforcement Learning (RL), specifically the DDPG \cite{DDPG}, which integrates the DQN \cite{DQN} and actor-critic \cite{actor-critic} methodologies to optimize in high-dimensional spaces. However, CDBTune requires extensive training time and benchmarking, resulting in long tuning durations. Ranking weighted Gaussian process ensemble (RGPE) \cite{Facilitating} is an ensemble model for BO-based optimization, which combines information about the workload and historical knowledge to optimize the process. However, RGPE faces challenges when optimizing high-dimensional data, presenting a similar problem to OtterTune, where the selection of parameters for optimization does not guarantee stable performance. Furthermore, RGPE experiences limitations in setting appropriate hyperparameter values for the model.

\begin{figure*}[t]
    \centering
    \includegraphics[width=1\linewidth]{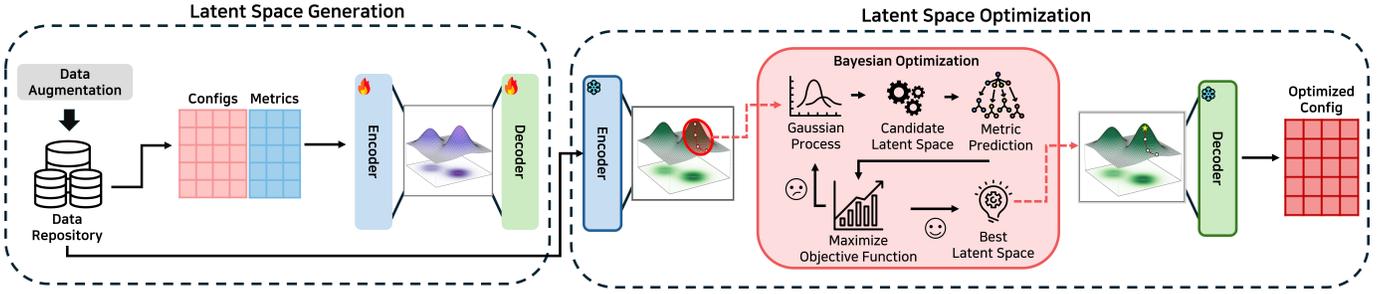}
    \caption{Overview Architecture of LatentTune}
    \label{fig:overview}
\end{figure*}

\section{METHOD}

Our proposed framework consists of several modules, as illustrated in Figure~\ref{fig:overview}.
To address the challenge of collecting large amounts of training data for database configuration tuning, our approach begins with data augmentation applied to a small set of measured configurations and their corresponding performance metrics. By leveraging both the original and augmented samples, we train a low-dimensional representation space that captures the essential relationships between configuration parameters and performance outcomes. The encoder compresses high-dimensional configurations into a compact latent space, while the decoder reconstructs them back into the original configuration space. Finally, Bayesian Optimization (BO) is performed in the latent space to efficiently search for high-performing configurations, and the optimal latent vectors are decoded into complete configuration sets for deployment. Further details of each module are provided in the following subsections.

\subsection{Data Augmentation}
ML-based methods require large amounts of training data to improve model accuracy. 
However, in the field of database tuning research, there are no available public datasets due to the significant influence of various hardware conditions and the specific workloads (sets of tasks or queries) each database system processes. Consequently, the training data generation must be conducted individually, however it requires considerable resources and time. To address this limitation, we employ Sample Augmentation process.

The Sample Augmentation module augments the configuration and metric data pairs. We define the original dataset as $\mathcal{D} = \left\{ \left( \textit{conf}_i, y_i \right) \right\}_{i=1}^{N}$, where each configuration $\textit{conf}_i \in \mathbb{R}^d$ is a $d$-dimensional space of database parameters, and $y_i \in \mathbb{R}^e$ is an $e$-dimensional vector of external performance metrics. Similarly, the augmented configuration and metric pairs are denoted by $\textit{conf}'_i \in \mathbb{R}^d$ and $y'_i \in \mathbb{R}^e$. The resulting augmented dataset is represented as $\hat{\mathcal{D}} = \left\{ \left(\textit{conf}'_i, y'_i \right) \right\}_{i=1}^{M}$.

In the configuration sampling phase, we employ LHS~\cite{LHS} to generate configuration data $conf'$. LHS is a stratified sampling technique that enables uniform and efficient exploration of multi-dimensional parameter spaces. It ensures that samples are evenly distributed within each dimension, thereby achieving a well-balanced coverage of the overall input space. This systematic approach allows for a more accurate reflection of the individual and combined impact of parameters on performance metrics~\cite{Facilitating, robotune, hunter, locat}. In contrast, conventional methods such as Simple Random Sampling~\cite{randomsampling} and Gaussian Sampling~\cite{gaussian} often suffer from poor coverage in high-dimensional spaces, particularly when parameter ranges are wide. These methods tend to produce clustered or sparse samples, leading to biased or inefficient exploration of the configuration space. 

Given these limitations, LHS enables more uniform and comprehensive coverage of the search space, making it
particularly suitable for database parameter tuning scenarios. Therefore, we adopt LHS for generating configuration samples in our framework.

The process of measuring metric values using benchmarking tools is time-consuming and highly resource-intensive. Thus, we utilized a prediction model to predict metric values for $conf'$. Specifically, we employ TabNet~\cite{TabNet}, which is suitable for tabular data and efficient for multi-target label prediction. TabNet employs a mask-based attention mechanism to effectively identify and prioritize important features among numerous columns, thereby enhancing predictive performance by focusing on highly influential attributes while suppressing less informative ones. For example, database configuration data usually contain numerous parameters and corresponding values organized in a tabular format. TabNet systematically identifies and learns the importance of each parameter, enabling accurate predictions of external performance metrics. 

TabNet is trained with the original dataset $\mathcal{D}$ and used to predict metric values $y'$ for the augmented configuration data $conf'$. The augmented dataset $\hat{\mathcal{D}}$ is then constructed by combining the LHS sampled configurations $conf'$ with the corresponding predicted values $y'$. The prediction accuracy of TabNet is evaluated in the Section \ref{Analysis of Sample Augmentation}.

\subsection{Latent Space Generation}
We leverage an AE~\cite{AE} to construct a latent space that captures the underlying relationships between database configurations and external metrics. During the latent space optimization (LSO) phase, the decoder of the trained AE is employed to reconstruct the optimized latent vectors into the original high-dimensional configuration space, enabling direct deployment of the resulting configurations.

\noindent\textbf{Training.}  
To generate an informative latent space, LatentTune trains the AE on both the original dataset $\mathcal{D}$ and the augmented dataset $\hat{\mathcal{D}}$. Each training sample is formed by concatenating a configuration and its corresponding performance metric, resulting in an input data $I \in \mathbb{R}^t$, where $t = d + e$, with $d$ and $e$ denoting the dimensions of configuration and metric, respectively. The reason for concatenating the configuration and metric is to inject workload information into the latent space. Since configurations are composed of arbitrary numerical values, it is difficult to extract meaningful semantic representations in the latent space using configuration data solely. Therefore, by concatenating the configuration and its corresponding performance metric as input, the latent space is able to capture the dependencies between parameters and performance. This approach also allows the latent space to implicitly encode information about diverse target workloads.

The AE is trained to minimize the reconstruction loss between the input and its decoded output. Specifically, it minimizes the following objective:
\begin{equation}
\mathcal{L}_{\text{AE}} = \frac{1}{N+M} \sum_{i=1}^{N+M} \left\| x_i - \mathcal{D}_{\theta}(\mathcal{E}_{\phi}(x_i)) \right\|^2
\end{equation}
Here, $x_i$ represents the input data (concatenated configuration and metric), $\mathcal{E}(\cdot)$ and $\mathcal{D}(\cdot)$ are denote the encoder and decoder, respectively. Once training is complete, the encoder compresses the high-dimensional input data into a lower-dimensional latent space, while the decoder reconstructs it back to the original input data space. The LSO process then performs optimization in this latent space, and the optimized latent vectors are decoded to generate final configuration candidates.

\noindent\textbf{Encoder.} $\mathcal{E}: \mathbb{R}^t \rightarrow \mathbb{R}^l$ maps the high-dimensional input data $I$ to a low-dimensional latent representation $z \in \mathbb{R}^l$, where $l \ll t$, preserving essential features while reducing dimensionality.

\noindent\textbf{Decoder.} $\mathcal{D}: \mathbb{R}^l \rightarrow \mathbb{R}^t$ reconstructs the latent space back into the original high-dimensional space. Specifically, the optimized latent vector $z^\ast$ obtained from the LSO step is reconstructed through the trained decoder to produce the final output, consisting of a configuration and its corresponding metric.

\subsection{Latent Space Optimization.}
The latent space $z \in [0,1]$ extracted via the LSG step is optimized by BO~\cite{BO}. BO is a probabilistic optimization approach designed to efficiently explore and exploit the search space of black-box functions, especially in scenarios where function evaluations are computationally expensive and no explicit functional form is available. In each iteration, BO aims to maximize the value of a black-box objective function by iteratively constructing a probabilistic surrogate model and selecting candidate points through an acquisition function that balances exploration and exploitation. 

In this process, we employ a Gaussian Process (GP)~\cite{GP} as a surrogate model in the latent space to capture the posterior distribution of the objective. Given a candidate latent vectors $z$, we use the Expected Improvement (EI) acquisition function, which quantifies the expected improvement over the best observed performance. Formally,
\begin{equation}
\small
\mathrm{EI}(z) = (\mu(z) - f(z^*) - \xi) \cdot \Phi(\gamma(z)) + \sigma(z) \cdot \phi(\gamma(z))
\label{eq:ei}
\end{equation}
where $\mu(z)$ and $\sigma(z)$ are the GP predicted mean and standard deviation, respectively, and $f(z^*)$ is the best observed performance so far. The normalized improvement $\gamma(z)$ is computed as:
\begin{equation}
\gamma(z) = \frac{\mu(z) - f(z^*) - \xi}{\sigma(z)}
\end{equation}
where $\xi$ is a small positive constant that balances exploration and exploitation.

The objective function is typically evaluated as performance (i.e., metrics) by a benchmarking tool. However, since benchmarking is a time-consuming process, we implemented a performance prediction model based on TabNet. It is important to note that this model does not share weights with the TabNet model used in the sample augmentation step.

TabNet is trained on the latent space $z$, generate through the LSG process, along with the corresponding metrics $y$ and $y'$ . In other words, the input to the prediction model is the latent space, and the output is the metric value. As TabNet’s predictions are multi-label, a score function is defined to convert these predict value into a scalar function, facilitating smoother comparisons. The objective of the optimization process is to maximize this score function.

For MySQL, the primary goal is to maximize \texttt{Throughput} while minimizing \texttt{Latency}. To reflect this objective, we define the score function as follows: 
\begin{equation}
\small
f_{\text{MySQL}}(x) = \frac{\operatorname{Throughput}(x)}{\operatorname{Latency}(x)}
\end{equation}

For RocksDB, multiple performance metrics must be optimized simultaneously. Specifically, given four performance metrics (\texttt{TIME}, \texttt{RATE}, \texttt{WAF}, and \texttt{SAF}), the details of each metric are provided in Section \ref{Experiment Setup}, Evaluation Metrics.

\noindent we define the score function as:
\begin{equation}
\small
f_{\text{RocksDB}}(x) = -w_0 \cdot \text{TIME} + w_1 \cdot\text{RATE} - w_2\cdot \text{WAF} - w_3 \cdot \text{SAF}
\end{equation}

\noindent where, \texttt{TIME}, \texttt{WAF}, and \texttt{SAF} are metrics that should be minimized, and \texttt{RATE} is a metric that should be maximized. $w_0, w_1, w_2,$ and $w_3$ represent the weights assigned to each of the four metrics.


\section{EXPERIMENT}
\subsection{Experiment Setup} \label{Experiment Setup}

\noindent{\textbf{Hardware.}} All MySQL performance measurements were performed on a server running the Ubuntu 20.04.6 operating system with an Intel® Core™ i7-11700 @ 2.52GHZ, 32 GB of RAM, and a 256 GB disk. All RocksDB performance measurements were performed with a server running the CentOS 7.9 operating system on an Intel® Core™ i5-8500 CPU @ 3.00 GHz, 32 GB of RAM, and a 256 GB disk. \\

\noindent{\textbf{Tuning Setting.}} We evaluated performance using MySQL v5.7.37 \cite{mysql} and Rocks-DB v6.25.0 \cite{rocksdb}. We select 138 tunable parameters in MySQL via OtterTune without knobs related to debugging, security, or path settings. The initial dataset consists of 1,000 samples, and it is augmented to 5,000 samples through the data augmentation process. A discussion of the optimal number of augmented samples is presented in 4.3 section. The experiment was repeated five times with different random seed values for each workload, and the average values were calculated. In the baseline experiment, the top-$k$ selected parameters for tuning were set to 5, 10, and 15, respectively, and the best-performing values were used for comparison. \\

\noindent{\textbf{Workload.}} We conducted experiments on four different workloads for both MySQL and RocksDB. The details of all workloads are shown in Table \ref{tab:mysql-table} and \ref{tab:rocksdb-table}. 
\textbf{YCSB}(Yahoo! Cloud Serving Bench-mark)} \cite{ycsb} is a benchmark tool offers several workloads, each defining the rate and pattern of various operations (load, insert, read, update, scan, etc.) on the database. It can be used to compare and evaluate the performance of various database systems, including MySQL. For RocksDB workloads, we experimented with four workloads with different percentages for read, write, and update. \\

\noindent{\textbf{Evaluation Metrics.}} 
The metrics we used to compare the performance improvement between the baseline and LatentTune are as follows. \textbf{MySQL} uses two evaluation metrics: \texttt{Throughput}, which represents the amount of query processing that can be completed within a given time, and \texttt{Latency}, which refers to the delay time required to process the queries. 

\begin{table}[t]
    \centering
    \caption{MySQL Workload Information}
    \label{tab:mysql-table}
    \small
    \setlength{\tabcolsep}{2pt} 
    \setlength{\extrarowheight}{0pt} 
    \begin{tabular}{cccccccc}
        \hline
        \textbf{MySQL} & \textbf{} & \textbf{} & \multicolumn{1}{l}{} & \multicolumn{1}{l}{} & & \multicolumn{1}{l}{} & \multicolumn{1}{l}{} \\ \hline
        \multicolumn{1}{c|}{\begin{tabular}[c]{@{}c@{}}Workload\\Index\end{tabular}} 
        & \multicolumn{1}{c|}{\begin{tabular}[c]{@{}c@{}}Scale\\Factor\end{tabular}} 
        & \multicolumn{1}{c|}{\begin{tabular}[c]{@{}c@{}}Data\\Size\end{tabular}} 
        & \multicolumn{1}{l|}{Read} & \multicolumn{1}{l|}{Insert} 
        & \multicolumn{1}{c|}{Scan} & \multicolumn{1}{l|}{Update} 
        & \multicolumn{1}{l}{\begin{tabular}[c]{@{}l@{}}Read\\Modify\\Write\end{tabular}} \\ \hline
        \multicolumn{1}{c|}{A} & \multicolumn{1}{c|}{\multirow{4}{*}{12000}} & \multicolumn{1}{c|}{\multirow{4}{*}{15GB}} 
        & \multicolumn{1}{c|}{\textbf{50\%}} & \multicolumn{1}{c|}{-} 
        & \multicolumn{1}{c|}{-} & \multicolumn{1}{c|}{\textbf{50\%}} & - \\
        \multicolumn{1}{c|}{B} & \multicolumn{1}{c|}{} & \multicolumn{1}{c|}{} & \multicolumn{1}{c|}{\textbf{95\%}} & \multicolumn{1}{c|}{-} 
        & \multicolumn{1}{c|}{-} & \multicolumn{1}{c|}{\textbf{5\%}} & - \\
        \multicolumn{1}{c|}{E} & \multicolumn{1}{c|}{} & \multicolumn{1}{c|}{} & \multicolumn{1}{c|}{-} & \multicolumn{1}{c|}{\textbf{5\%}} 
        & \multicolumn{1}{c|}{\textbf{95\%}} & \multicolumn{1}{c|}{-} & - \\
        \multicolumn{1}{c|}{F} & \multicolumn{1}{c|}{} & \multicolumn{1}{c|}{} & \multicolumn{1}{c|}{\textbf{50\%}} & \multicolumn{1}{c|}{-} 
        & \multicolumn{1}{c|}{-} & \multicolumn{1}{c|}{-} & \textbf{50\%} \\ \hline
    \end{tabular}
\end{table}

\begin{table}[t]
    \centering
    \caption{RocksDB Workload Information}
    \label{tab:rocksdb-table}
    \small
    \setlength{\tabcolsep}{2pt}
    \setlength{\extrarowheight}{2pt}
    \begin{tabular}{cccccc}
        \hline
        \textbf{RocksDB}                       & \textbf{}                          & \textbf{}                           &                                    &                \\ \hline                                                                                                                                                    
         \multicolumn{1}{c|}{\begin{tabular}[c]{@{}c@{}}Workload \\ Index\end{tabular}} & 
        \multicolumn{1}{c|}{\begin{tabular}[c]{@{}c@{}}Value Size\end{tabular}} & 
        \multicolumn{1}{c|}{\begin{tabular}[c]{@{}c@{}}\# of Entry\end{tabular}} & 
        \multicolumn{1}{c|}{READ} & 
        \multicolumn{1}{c|}{WRITE} & 
        UPDATE \\ \hline
        \multicolumn{1}{c|}{R90W10}              & \multicolumn{1}{c|}{\multirow{4}{*}{16384}} & \multicolumn{1}{c|}{\multirow{4}{*}{65472}} & \multicolumn{1}{c|}{90\%} & \multicolumn{1}{c|}{10\%}  & \multirow{3}{*}{X} \\ 
        \multicolumn{1}{c|}{R50W50}              & \multicolumn{1}{c|}{}                       & \multicolumn{1}{c|}{}                       & \multicolumn{1}{c|}{50\%} & \multicolumn{1}{c|}{50\%}  &                    \\ 
        \multicolumn{1}{c|}{R10W90}             & \multicolumn{1}{c|}{}                       & \multicolumn{1}{c|}{}                       & \multicolumn{1}{c|}{10\%} & \multicolumn{1}{c|}{90\%}  &                    \\ 
        \multicolumn{1}{c|}{UPDATE}             & \multicolumn{1}{c|}{}                       & \multicolumn{1}{c|}{}                       & \multicolumn{1}{c|}{-}    & \multicolumn{1}{c|}{-}     & O                  \\ \hline
    \end{tabular}
\end{table}

Higher throughput and lower latency indicate better performance, the latency being measured at the 99th percentile. \textbf{RocksDB} performance metrics include \texttt{TIME}, which measures the time taken to complete database operations for a given workload; \texttt{RATE}, which measures the number of operations processed per second; Write Amplification Factor (\texttt{WAF}) which indicates the write amplification ratio of the workload for a specific configuration; and Space Amplification Factor (\texttt{SAF}), which represents the space amplification ratio during the data storage process. Good performance is defined by lower TIME, WAF, and SAF values, and a higher RATE value. To compare the performance improvement rates of the four metrics in RocksDB, we calculated the following score:

{
\small
\begin{equation}
\begin{split}
\text{\textit{SCORE}} = & \log\left(\frac{\text{TIME}_\text{default}}{\text{TIME}}\right) 
+ \log\left(\frac{\text{RATE}}{\text{RATE}_\text{default}}\right) \\
& + \log\left(\frac{\text{WAF}_\text{default}}{\text{WAF}}\right) 
+ \log\left(\frac{\text{SAF}_\text{default}}{\text{SAF}}\right)
\end{split}
\end{equation}
}

This score represents the performance improvement of the optimized parameters compared to the default settings, with the logarithms of individual improvement rates added for easier comparison. \\

\begin{figure}[t]
    \centering
    \includegraphics[height=0.60\textheight]{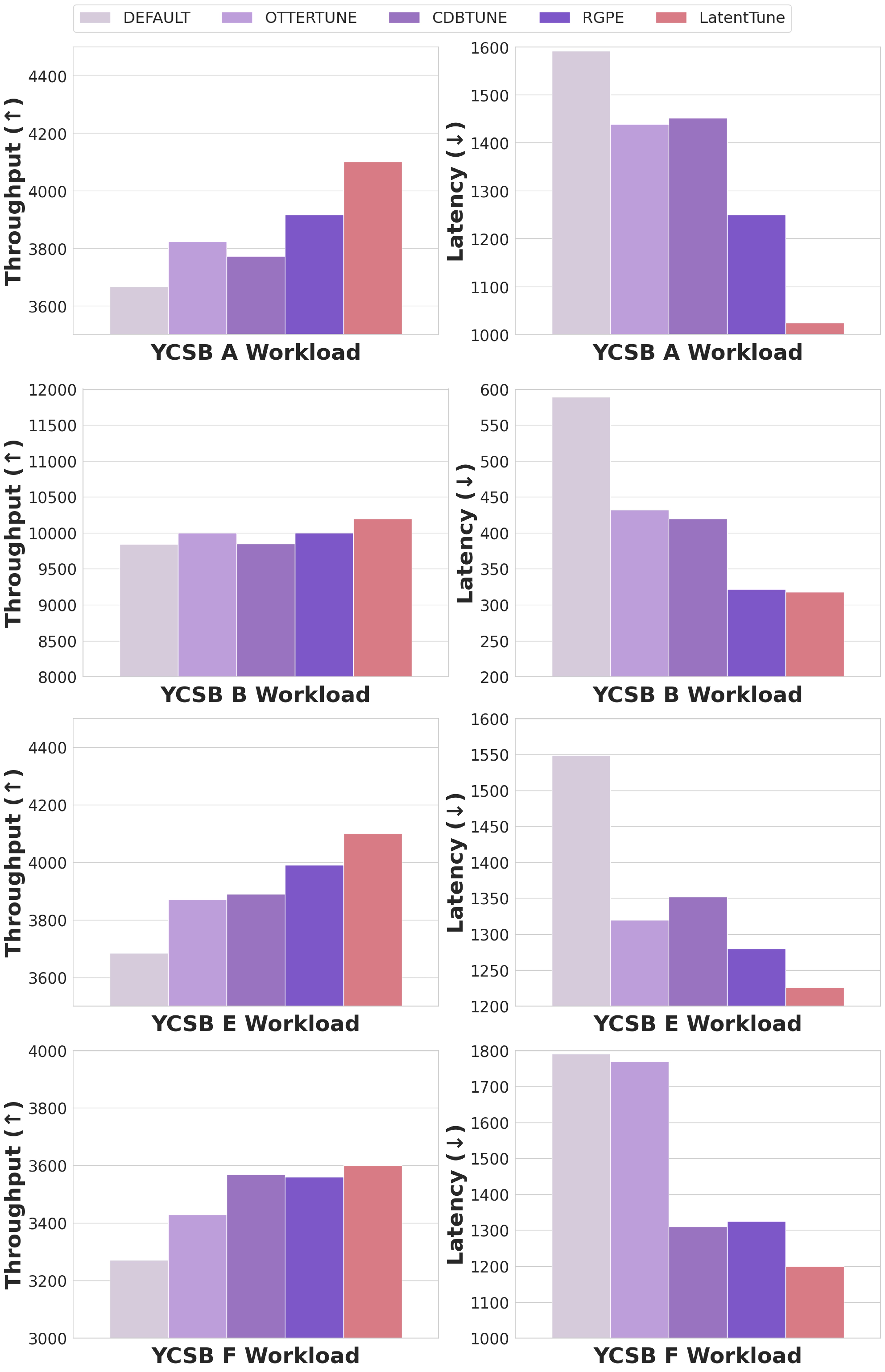} 
    \caption{Performance improvement for MySQL YCSB A,B,E,F workload}
    \label{fig:Mysql_results}
\end{figure}

\noindent{\textbf{Implementation Details.}} Our model projected to 32 dimensions for MySQL, and 16 for RocksDB. We set the same 300 iterations for all BO processes in our experiments. 

\subsection{Efficiency Comparison}
We compared LatentTune with OtterTune, CDBTune, and RGPE on four workloads for MySQL and RocksDB. Each of the three baselines tuned a selection of the top parameters in terms of impact on database performance, and each model included a workload mapping process.

Figure \ref{fig:Mysql_results} presents the experimental results on MySQL. The results demonstrate that LatentTune achieved the highest improvement rates in both throughput and latency across all workloads. For workload B, only slight performance improvements are observed across both LatentTune and the baseline methods. This is because workload B is a read-heavy workload composed of lightweight operations, which inherently place minimal stress on the system. As a result, changes in database configurations have limited impact on throughput. However, latency can still be significantly affected by parameters related to concurrency control and I/O scheduling, leading to noticeable latency improvements with LatentTune.

Figure \ref{fig:RocksDB_results} shows the experimental results on RocksDB. To facilitate a clear comparison of performance improvements, the enhanced performance over the default configuration is represented using the score function. For all four workloads, LatentTune demonstrates improved performance with a maximum score of 9.19 and this result indicates that the LatentTune outperforms the baseline model.

Furthermore, we observed performance improvements across all four metrics compared to the baselines, enabling balanced optimization across multiple aspects of the database. The experimental results indicate that optimizing with information from all parameters is more effective than tuning only the most important ones, as the latter approach ignores the impact of the unselected parameters. Moreover, LatentTune demonstrates the ability to tune various workloads without a workload mapping process, unlike the three baselines. This is because the target workload information is injected into the low-dimensional space.

\subsection{Analysis of Sample Augmentation} \label{Analysis of Sample Augmentation}
\subsubsection{Performance comparison based on number of data}
To demonstrate the efficiency of data augmentation in the ML-based tuning process, we compared the AE’s reconstruction loss, and LatentTune’s performance improvement rate based on the amount of data.
Figure \ref{fig:recon_loss} shows the AE's reconstruction loss based on the number of training data. Reconstruction loss refers to the deviation between the AE's input and output data by the decoder. In other words, a lower loss indicates higher-quality reconstruction performance. The experimental results show that the reconstruction loss of the AE gradually decreases as the number of training data increases, and the lowest loss is observed when the model is trained with 5,000 datasets. As the number of data points exceeds 5,500, increased reconstruction loss is observed because the information in the data is limited, and the AE’s ability to learn is also constrained. This means that even with data augmentation, there is an upper limit to the amount of relevant information that the model can learn, and AE performance can be expected to decrease if that threshold is exceeded. Moreover, when a large amount of data is augmented, the relative proportion of the original data decreases. As a result, the accuracy of the information in the latent space compressed by the AE diminishes, leading to a subsequent increase in reconstruction loss. This consequently leads to a decline in the performance enhancement achieved by LatentTune.

\begin{figure}[t]
    \centering
    \includegraphics[width=0.5\textwidth]{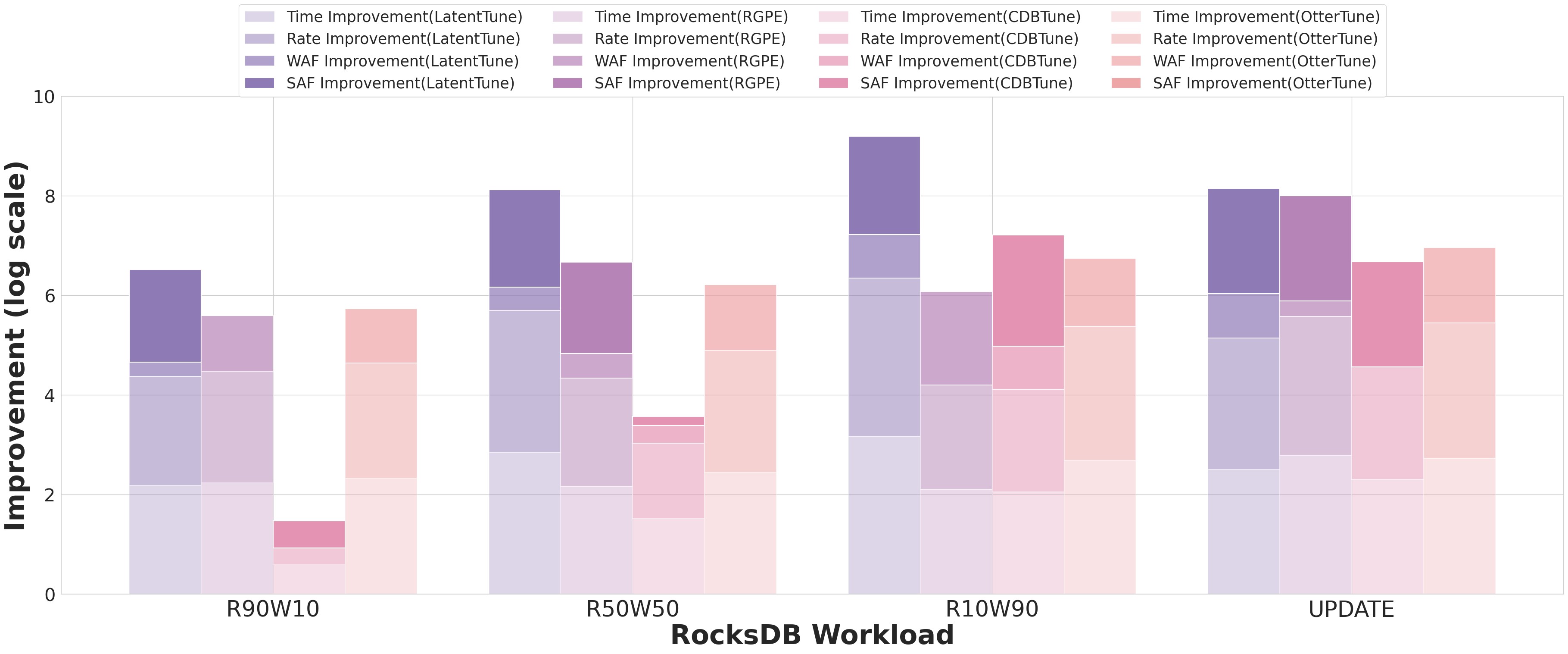}
    \caption{Performance improvement for RocksDB workloads}
    \label{fig:RocksDB_results}
\end{figure}

Figure \ref{fig:performance_improvement} shows LatentTune’s throughput improvement rate based on the number of training data. Similar to the lowest reconstruction loss, the highest improvement rate is observed when the number of training data is 5,000.

In conclusion, the AE exhibited optimal performance when the data were augmented for a total of 5,000 datasets. Furthermore, the latent space generated from a well-trained AE successfully compresses information regarding the database parameters and target workload. These results also confirm that LatentTune maximizes database performance by optimizing the latent space. Accordingly, we set the number of augmented data samples to 5,000 when configuring our model experiments.

\begin{table*}[t!]
\centering
\caption{Metric Prediction Model Performance for MySQL (YCSB E, YCSB F) and RocksDB (R10W90, UPDATE) in Sample Augmentation.}
\label{tab:aug-prediction}
\setlength{\tabcolsep}{2pt} 
\renewcommand{\arraystretch}{1.2} 
\resizebox{0.8\textwidth}{!}{
\begin{tabular}{l|c|c|c|c| |c|c|c|c|c|c|c|c}
\hline
\multicolumn{5}{c||}{\textbf{YCSB E}} & \multicolumn{8}{c}{\textbf{R10W90}} \\ \hline
\multirow{2}{*}{Model} & \multicolumn{2}{c|}{Throughput} & \multicolumn{2}{c||}{Latency} 
                      & \multicolumn{2}{c|}{TIME} & \multicolumn{2}{c|}{RATE} & \multicolumn{2}{c|}{WAF} & \multicolumn{2}{c}{SAF} \\ \cline{2-13}
 & R2(↑) & MSE(↓) & R2(↑) & MSE(↓)
 & R2(↑) & MSE(↓) & R2(↑) & MSE(↓) & R2(↑) & MSE(↓) & R2(↑) & MSE(↓) \\ \hline
MLP           & 0.08 & 0.810 & 0.09 & 0.780 & 0.45 & 0.0060 & 0.31 & 0.00900 & 0.40 & 0.0030 & 0.45 & 0.06000 \\
Decision Tree & 0.84 & 0.140 & 0.83 & 0.144 & 0.83 & 0.0010 & 0.87 & 0.00100 & 0.50 & 0.0020 & 0.99 & 0.00005 \\
Random Forest & 0.99 & 0.006 & 0.96 & 0.026 & 0.85 & 0.0007 & 0.93 & 0.00090 & 0.76 & 0.0010 & 0.99 & 0.00002 \\
XGBoost       & \textbf{0.99} & \textbf{0.001} & \textbf{0.97} & \textbf{0.022} & 0.88 & 0.0005 & 0.93 & 0.00090 & 0.86 & 0.0007 & 0.99 & 0.00003 \\
CatBoost      & 0.98 & 0.016 & 0.95 & 0.033 & 0.89 & 0.0005 & 0.92 & 0.00080 & 0.87 & 0.0007 & 0.99 & 0.00007 \\
TabNet        & 0.98 & 0.010 & 0.95 & 0.035 & \textbf{0.90} & \textbf{0.0001} & \textbf{0.95} & \textbf{0.00003} & \textbf{0.87} & \textbf{0.0003} & \textbf{0.99} & \textbf{0.00001} \\ \hline

\multicolumn{5}{c||}{\textbf{YCSB F}} & \multicolumn{8}{c}{\textbf{UPDATE}} \\ \hline
\multirow{2}{*}{Model} & \multicolumn{2}{c|}{Throughput} & \multicolumn{2}{c||}{Latency}
                      & \multicolumn{2}{c|}{TIME} & \multicolumn{2}{c|}{RATE} & \multicolumn{2}{c|}{WAF} & \multicolumn{2}{c}{SAF} \\ \cline{2-13}
 & R2(↑) & MSE(↓) & R2(↑) & MSE(↓)
 & R2(↑) & MSE(↓) & R2(↑) & MSE(↓) & R2(↑) & MSE(↓) & R2(↑) & MSE(↓) \\ \hline
MLP           & 0.17 & 0.871 & 0.16 & 0.900 & 0.46 & 0.0030 & 0.35 & 0.007 & 0.51 & 0.030 & 0.46 & 0.0500 \\
Decision Tree & 0.98 & 0.018 & 0.95 & 0.045 & 0.80 & 0.0010 & 0.76 & 0.002 & 0.29 & 0.020 & 0.98 & 0.0004 \\
Random Forest & \textbf{0.98} & \textbf{0.012} & \textbf{0.97} & \textbf{0.022} & 0.90 & 0.0005 & 0.87 & 0.001 & 0.68 & 0.003 & 0.99 & 0.0005 \\
XGBoost       & 0.98 & 0.017 & 0.97 & 0.023 & 0.92 & 0.0004 & 0.87 & 0.001 & 0.75 & 0.003 & 0.99 & 0.0006 \\
CatBoost      & 0.97 & 0.021 & 0.96 & 0.037 & 0.92 & 0.0005 & 0.85 & 0.005 & 0.80 & 0.002 & 0.99 & 0.0005 \\
TabNet        & 0.96 & 0.043 & 0.93 & 0.070 & \textbf{0.94} & \textbf{0.0002} & \textbf{0.87} & \textbf{0.001} & \textbf{0.83} & \textbf{0.002} & \textbf{0.99} & \textbf{0.0004} \\ \hline
\end{tabular}
}
\end{table*}

 \begin{figure}[!t]
    \centering
    \begin{minipage}[t]{0.48\linewidth}
        \centering
        \includegraphics[width=\linewidth]{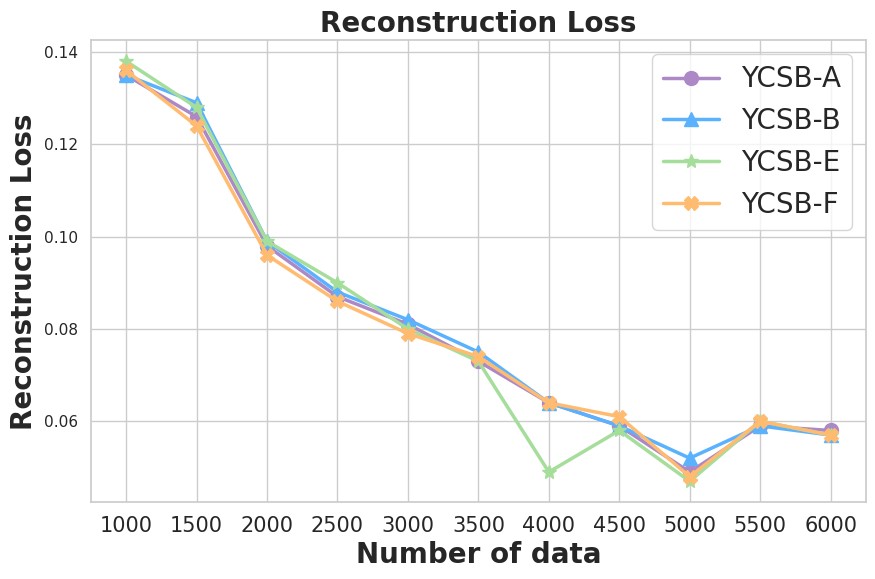}
        \caption{Reconstruction Loss with number of data.}
        \label{fig:recon_loss}
    \end{minipage}
    \hfill
    \begin{minipage}[t]{0.48\linewidth}
        \centering
        \includegraphics[width=\linewidth]{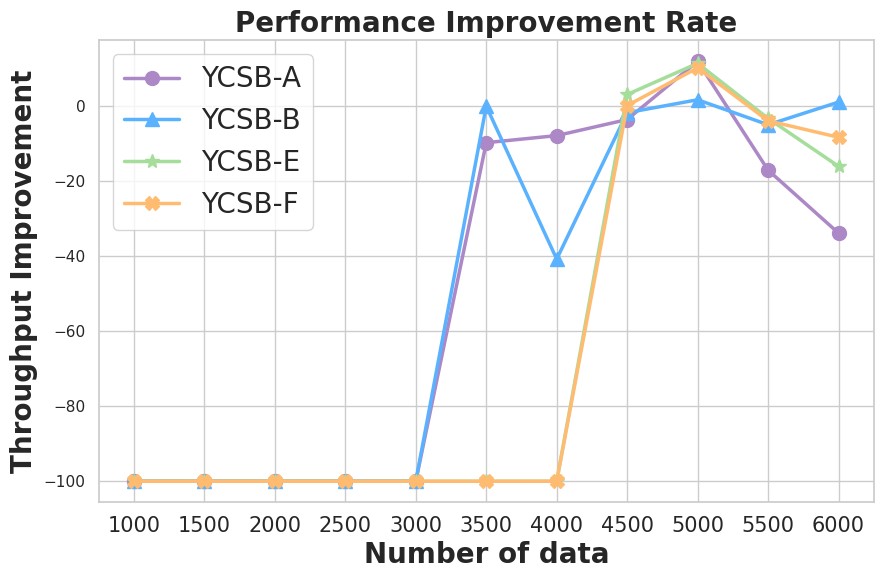}
        \caption{Performance Improvement with number of data.}
        \label{fig:performance_improvement}
    \end{minipage}
\end{figure}

\subsubsection{Prediction models for sample augmentation}
We compared the accuracy of the prediction model used in benchmarking stage during the sample augmentation process. We selected five models for comparison: basic multilayer perceptron (MLP)\cite{MLP}, two tree-based models\cite{decision,rf}, and two models with boosting techniques\cite{xg,cat}.

As shown in Table \ref{tab:aug-prediction}, Random Forest or XGBoost achieved the highest accuracy in MySQL with fewer prediction labels. However, as shown on the right side of Table \ref{tab:aug-prediction}, the results for RocksDB demonstrate that TabNet achieved superior accuracy with four labels. Although random forest showed good accuracy for MySQL, it failed to achieve consistent prediction accuracy across all metrics when predicting performance for RocksDB. The reason for this is that TabNet dynamically allocates important features for each label when making predictions for multi-label tasks. Therefore, we selected TabNet for sample augmentation due to its consistently high prediction accuracy across all labels, even with multi-label data.

\subsection{Analysis of Latent Space}
\subsubsection{Efficiency of Latent Space Optimization}
To evaluate the effectiveness of the proposed latent space optimization method, we analyzed the latent vectors for both optimized and non-optimized configurations in MySQL under two different workloads, YCSB A and YCSB F. 

As shown in Figure \ref{fig:latent_output}, the contour density plots illustrate that the latent vectors of non-optimized configurations (red) and optimized configurations (blue) form clearly distinct clusters in both workloads. This separation demonstrates that latent space optimization effectively captures and preserves meaningful structures from high-dimensional parameter spaces, ensuring that optimized configurations are well-represented within specific regions of the latent space. Furthermore, the clear distinction between the two clusters indicates the efficiency of our latent space optimization approach.
To further validate this, we reconstructed the latent space representations into the original high-dimensional space using the decoder of a trained AE and evaluated their actual performance. Figure \ref{fig:metrics_output} shows the contour density plots of the throughput-latency relationship for both workloads. The optimized configurations (blue) exhibit high throughput and low latency, aligning with optimal performance regions. In contrast, the non-optimized configurations (red) are concentrated in suboptimal regions characterized by lower throughput and higher latency. The clear distinction in contour density confirms that our method successfully optimizes configurations, and this pattern is consistently observed across different workloads.

Figure \ref{fig:metrics_box_AF_vertical} presents the inter-quartile range (IQR) of throughput and latency, respectively, as a detailed analysis of the performance metrics shown in Figure \ref{fig:metrics_output}. The results further confirm that optimized configurations achieve significantly higher throughput and lower latency compared to non-optimized configurations in both workloads. In particular, optimized configurations (blue) exhibit lower variance and no outliers, demonstrating that the optimization process improves both performance and stability.

\begin{figure}[t]  
    \begin{minipage}{\columnwidth}  
        \centering
        \includegraphics[width=0.9\columnwidth]{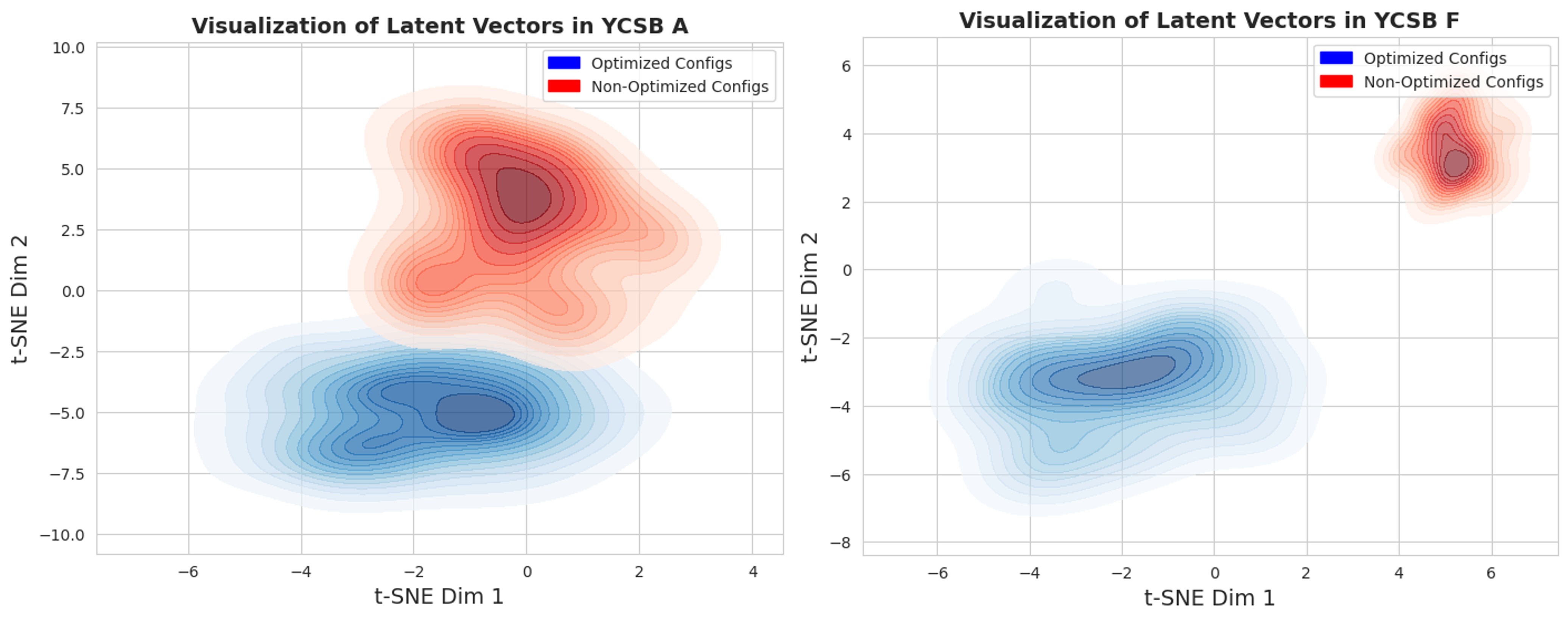}  
        \caption{Latent vectors density on YCSB A and YCSB F workload.}
        \label{fig:latent_output}
    \end{minipage}
    \vspace{-15pt}
\end{figure}

\begin{figure}[t]  
    \begin{minipage}{\columnwidth}  
        \centering
        \includegraphics[width=0.9\columnwidth]{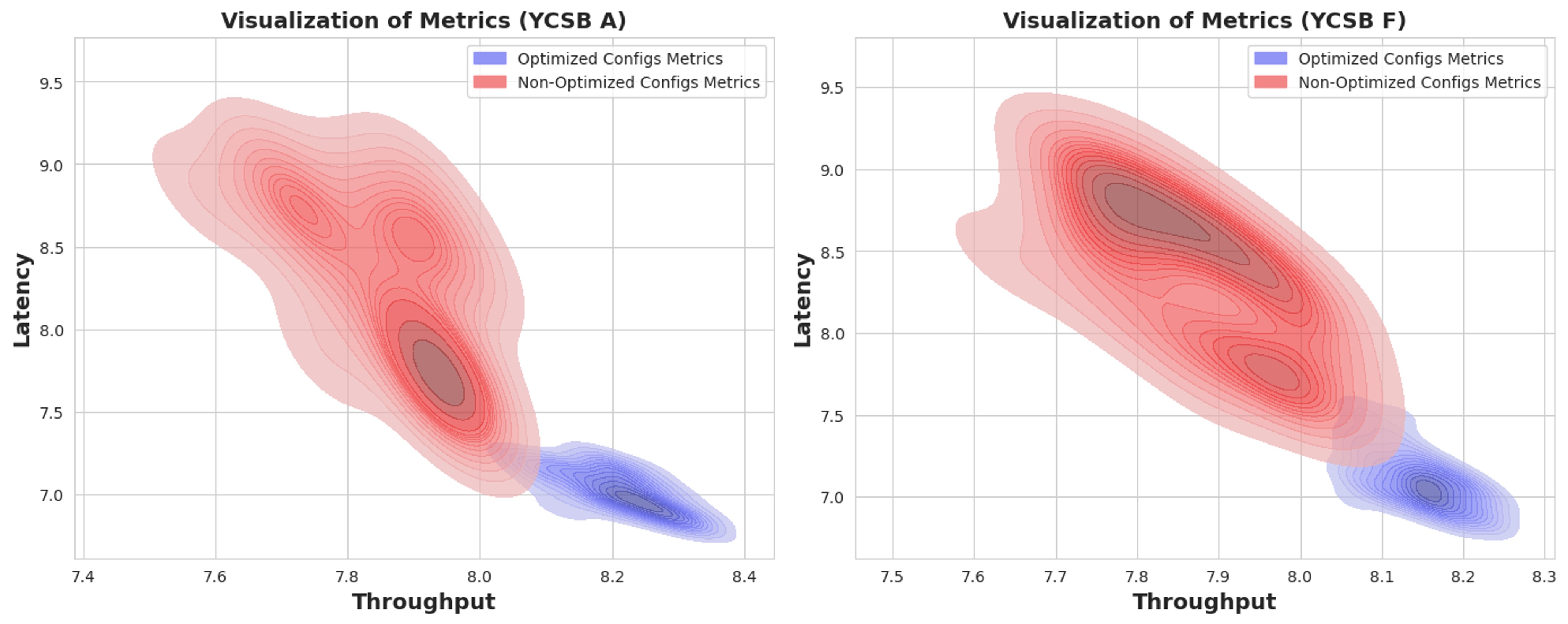}  
        \caption{Metrics density on YCSB A and YCSB F workload.}
        \label{fig:metrics_output}
    \end{minipage}
\end{figure}

\subsubsection{Prediction models for objective function in LSO}

In the LSO stage, instead of performing actual benchmarking in the objective function, we replaced the metric prediction model of the optimized configuration.We compare the accuracy of the prediction model for the optimized latent space, with the same settings as the experiment in Table \ref{tab:aug-prediction}.

\begin{table*}[t]
\centering
\caption{Comparison of Metric Prediction Model Performance for MySQL (YCSB E, YCSB F) and RocksDB (R90W10, R10W90) in BO.}
\label{tab:BO-predict}
\setlength{\tabcolsep}{2pt} 
\renewcommand{\arraystretch}{1.2} 
\resizebox{0.8\textwidth}{!}{
\begin{tabular}{l|c|c|c|c||c|c|c|c|c|c|c|c}
\hline
\multicolumn{5}{c||}{\textbf{YCSB E}} & \multicolumn{8}{c}{\textbf{R90W10}} \\ \hline
\multirow{2}{*}{Model} & \multicolumn{2}{c|}{Throughput} & \multicolumn{2}{c||}{Latency} 
                      & \multicolumn{2}{c|}{TIME} & \multicolumn{2}{c|}{RATE} & \multicolumn{2}{c|}{WAF} & \multicolumn{2}{c}{SAF} \\ \cline{2-13}
 & R2(↑) & MSE(↓) & R2(↑) & MSE(↓)
 & R2(↑) & MSE(↓) & R2(↑) & MSE(↓) & R2(↑) & MSE(↓) & R2(↑) & MSE(↓) \\ \hline
MLP           & 0.28 & 0.850 & 0.24 & 0.600 & 0.41 & 0.010 & 0.48 & 0.010 & 0.48 & 0.006 & 0.78 & 0.020 \\
Decision Tree & -0.47 & 0.080 & -0.51 & 0.040 & -0.30 & 0.040 & -0.26 & 0.030 & -0.29 & 0.170 & 0.22 & 0.080 \\
Random Forest & 0.18 & 0.040 & 0.16 & 0.020 & 0.37 & 0.010 & 0.43 & 0.010 & 0.41 & 0.007 & 0.68 & 0.003 \\
XGBoost       & 0.16 & 0.050 & 0.14 & 0.020 & 0.44 & 0.010 & 0.50 & 0.010 & 0.59 & 0.005 & 0.74 & 0.020 \\
CatBoost      & 0.45 & 0.030 & 0.42 & 0.010 & 0.49 & 0.010 & 0.54 & 0.010 & 0.61 & 0.005 & 0.79 & 0.020 \\
TabNet        & \textbf{0.85} & \textbf{0.008} & \textbf{0.79} & \textbf{0.005} & \textbf{0.80} & \textbf{0.004} & \textbf{0.72} & \textbf{0.005} & \textbf{0.75} & \textbf{0.003} & \textbf{0.96} & \textbf{0.003} \\ \hline

\multicolumn{5}{c||}{\textbf{YCSB F}} & \multicolumn{8}{c}{\textbf{R10W90}} \\ \hline
\multirow{2}{*}{Model} & \multicolumn{2}{c|}{Throughput} & \multicolumn{2}{c||}{Latency}
                      & \multicolumn{2}{c|}{TIME} & \multicolumn{2}{c|}{RATE} & \multicolumn{2}{c|}{WAF} & \multicolumn{2}{c}{SAF} \\ \cline{2-13}
 & R2(↑) & MSE(↓) & R2(↑) & MSE(↓)
 & R2(↑) & MSE(↓) & R2(↑) & MSE(↓) & R2(↑) & MSE(↓) & R2(↑) & MSE(↓) \\ \hline
MLP           & 0.024 & 0.070 & 0.023 & 0.030 & 0.65 & 0.002 & 0.61 & 0.0070 & 0.18 & 0.002 & 0.82 & 0.010 \\
Decision Tree & -0.34 & 0.120 & -0.39 & 0.400 & 0.09 & 0.005 & 0.28 & 0.0080 & -0.83 & 0.006 & 0.34 & 0.070 \\
Random Forest & 0.21 & 0.040 & 0.20 & 0.020 & 0.58 & 0.002 & 0.69 & 0.0030 & 0.13 & 0.003 & 0.74 & 0.020 \\
XGBoost       & 0.29 & 0.040 & 0.23 & 0.020 & 0.62 & 0.002 & 0.77 & 0.0080 & 0.16 & 0.002 & 0.78 & 0.002 \\
CatBoost      & 0.42 & 0.030 & 0.40 & 0.010 & 0.47 & 0.001 & 0.81 & 0.0020 & 0.22 & 0.002 & 0.73 & 0.020 \\
TabNet        & \textbf{0.92} & \textbf{0.004} & \textbf{0.86} & \textbf{0.004} & \textbf{0.72} & \textbf{0.001} & \textbf{0.93} & \textbf{0.0007} & \textbf{0.78} & \textbf{0.001} & \textbf{0.79} & \textbf{0.002} \\ \hline
\end{tabular}
}
\end{table*}

\begin{figure}[t]
    \centering
    \includegraphics[height=4cm]{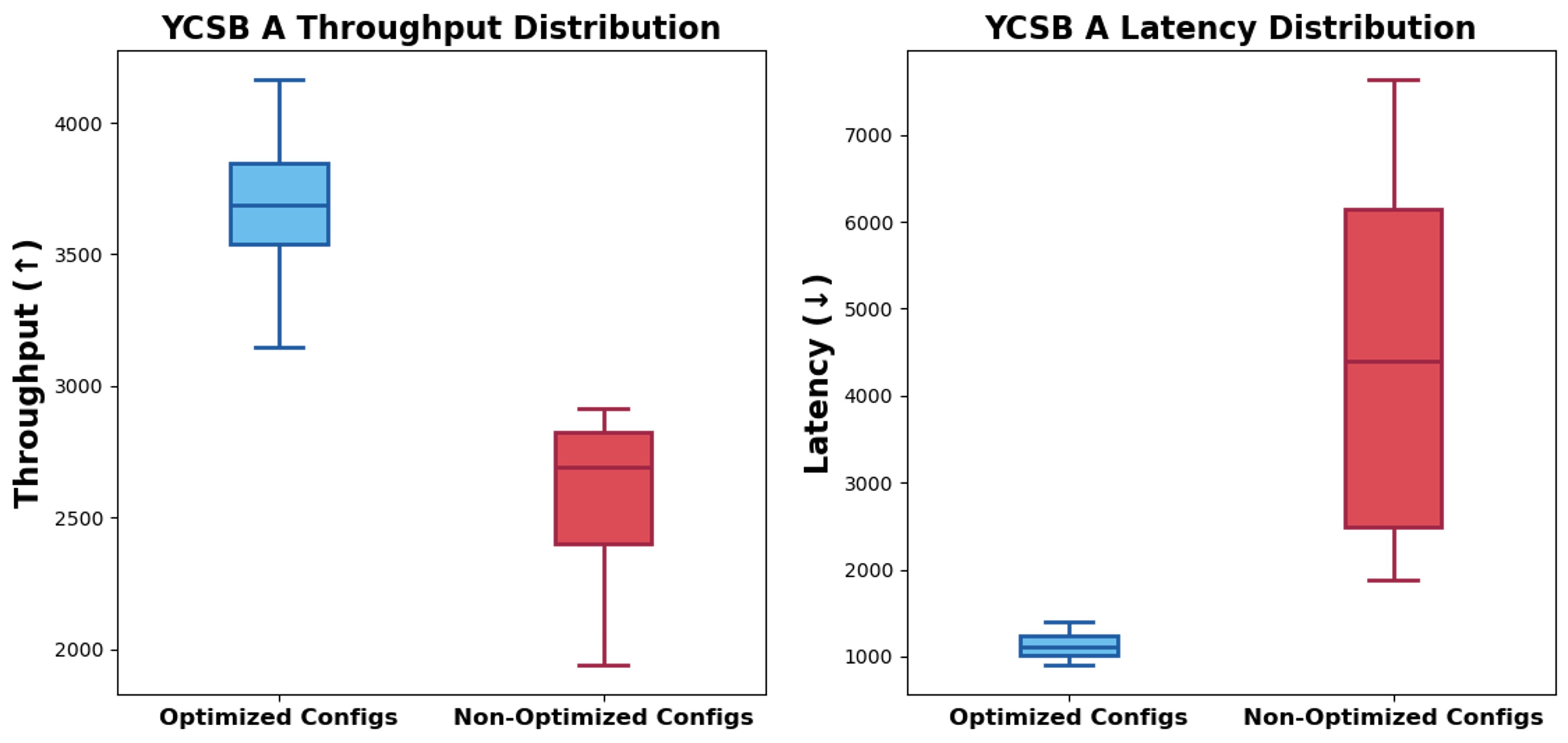}
    \vspace{0.5em}
    \includegraphics[height=4cm]{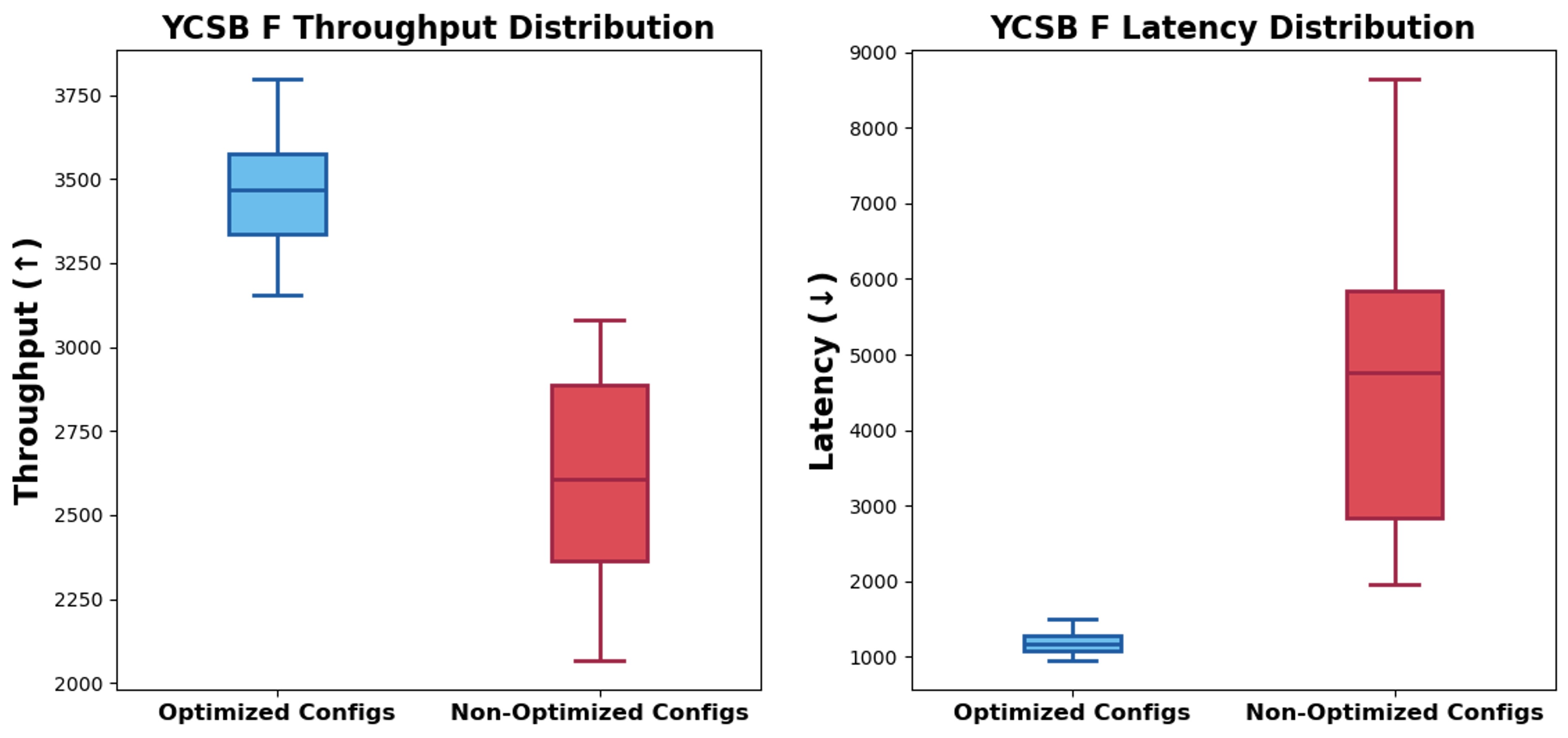}
    \caption{Metrics distribution on YCSB A (top) and YCSB F (bottom) workloads.}
    \label{fig:metrics_box_AF_vertical}
\end{figure}

As shown in Table \ref{tab:BO-predict}, the other prediction models either demonstrated poor performance or failed to predict accurately across all labels. However, TabNet achieved the highest accuracy in predicting metrics for both MySQL and RocksDB, and demonstrated consistent accuracy across all labels. Input data of the prediction model is in the form of tabular data, where the rows contain information about configurations and metrics, and the columns represent compressed key-features of the original data. As TabNet is well-suited for tabular data prediction, it outperforms the five comparison models and demonstrates equivalent performance across all labels.

\subsection{Analysis of Tuning Overhead and Efficiency}
Table \ref{tab:time_cost} compares the tuning time across two MySQL workloads.
Although RelTune did not achieve the shortest tuning time, it achieved the best overall optimization performance.
The relatively higher time cost of RelTune arises from the additional training required to construct the latent space via the autoencoder.
However, this cost remains moderate since the objective function in Bayesian Optimization (BO) is replaced with a lightweight prediction model.
This result highlights a trade-off between tuning efficiency and optimization performance—RelTune slightly increases the preprocessing time but significantly accelerates convergence toward optimal configurations.
Furthermore, RelTune exhibits superior data efficiency, achieving better results with fewer benchmark samples.
In practical settings, such a modest overhead is acceptable, as the gain in tuning stability and reproducibility outweighs the additional computation.

\subsection{Ablation Study}
We compared the effect of each module of our design on different workloads. Our experiments are conducted using various combinations of Sample Augmentation and LSG. All combinations are optimized using BO, with the same number of iterations set for the optimization process.

Table \ref{tab:ablation_combined} shows the results of the ablation experiments conducted on MySQL. Performance improvement for each combination demonstrates that the design of LatentTune with its modules is effective across all workloads. LatentTune leads to best performance improvement. Overall, combinations without sample augmentation and LSG showed very low performance, as BO in high-dimensional space leads to the curse of dimensionality.

This curse of dimensionality hinders the learning of the surrogate model in BO and significantly reduces the efficiency of exploration. For this reason, the first combination shows performance that is inferior to the default in all workloads. 

\begin{table}[t]
\caption{Time cost (minutes) to complete tuning across workloads. Lower is better.}
\label{tab:time_cost}
\centering
\small
\renewcommand{\arraystretch}{1.5}

\begin{tabular*}{\columnwidth}{@{\extracolsep{\fill}} l cccc @{}}
\toprule
\textbf{Method} & \textbf{YCSB A} & \textbf{YCSB B} & \textbf{YCSB E} & \textbf{YCSB F} \\
\midrule
OtterTune & \textbf{48.52}  & \textbf{50.16} & \textbf{49.70} & \textbf{48.24} \\
CDBTune   & 362.21 & 370.15 & 381.54  & 421.85  \\
RGPE      & 62.13 & 68.46  & 67.51 & 64.24  \\
LatentTune & 57.23 & 60.43 & 59.1  & 58.53  \\
\bottomrule
\end{tabular*}
\end{table}

When only the augmentation module is applied, the overall increase in the number of data achieve the surrogate model train, resulting in slightly improved performance compared to the first combination. However, due to the curse of dimensionality, the performance remains similar to the default.

When only the LSG module is applied, the reduction in dimensionality addresses the cure of dimensionality. However, since training an autoencoder also requires a large amount of data, the latent space is not well-compressed with the limited data without augmentation. In other words, it fails to effectively compress important
information from the data. This, in turn, degrades the performance of the prediction model used in BO to replace the objective function, leading to the poorest performance of the optimized configuration among all combinations.

\section{Conclusions}
In this paper, we addressed three limitations of ML-based database parameter tuning: the need for large training datasets, the tendency to tune only a subset of parameters due to high dimensionality, and reliance on similar workloads rather than the actual target workload. To overcome these challenges, we proposed LatentTune, a latent space-based tuning framework that dimensionality reduction with data augmentation. This allows for effective training with small datasets and enables full-parameter optimization. By injecting workload-specific information into the latent space, LatentTune also facilitates better adaptation to diverse workloads.

\begin{table}[t]
\renewcommand{\arraystretch}{1.1}
\caption{Ablation Study on MySQL with YCSB A,B and E,F.
"Aug" refers to data augmentation, and "LSG" denotes latent space generation.}
\label{tab:ablation_combined}
\centering
\small
\begin{minipage}{0.95\linewidth}
\centering
\textbf{(a) YCSB A and YCSB B} \vspace{0.3em}

\resizebox{\linewidth}{!}{
\begin{tabular}{c|c|cc|cc}
    \hline
    \multicolumn{2}{c|}{\textbf{Module}} & \multicolumn{2}{c|}{\textbf{YCSB A}} & \multicolumn{2}{c}{\textbf{YCSB B}} \\ \hline
    Aug & LSG & Throughput & Latency & Throughput & Latency \\ \hline
    X & X & 3651.22 & 1550 & 7448.72 & 714 \\
    O & X & 3689.79 & 1281 & 9952.72 & 611 \\
    X & O & 3314.65 & 1443 & 6570.68 & 903 \\
    \rowcolor{violet!10}
    O & O & \textbf{4100.34} & \textbf{1024} & \textbf{10000.00} & \textbf{318} \\ \hline
    \rowcolor{gray!10}
    \multicolumn{2}{c|}{Default} & 3666.90 & 1591 & 9845.56 & 589 \\ \hline
\end{tabular}
}
\end{minipage}

\vspace{1em} 

\begin{minipage}{0.95\linewidth}
\centering
\textbf{(b) YCSB E and YCSB F} \vspace{0.3em}

\resizebox{\linewidth}{!}{
\begin{tabular}{c|c|cc|cc}
    \hline
    \multicolumn{2}{c|}{\textbf{Module}} & \multicolumn{2}{c|}{\textbf{YCSB E}} & \multicolumn{2}{c}{\textbf{YCSB F}} \\ \hline
    Aug & LSG & Throughput & Latency & Throughput & Latency \\ \hline
    X & X & 3570.29 & 1588 & 3049.66 & 1220 \\
    O & X & 3996.52 & 1261 & 2934.99 & 1476 \\
    X & O & 3439.19 & 1782 & 2196.67 & 7001 \\
    \rowcolor{violet!10}
    O & O & \textbf{4100.45} & \textbf{1226} & \textbf{3600.76} & \textbf{1200} \\ \hline
    \rowcolor{gray!10}
    \multicolumn{2}{c|}{Default} & 3685.17 & 1549 & 3271.47 & 1791 \\ \hline
\end{tabular}
}
\end{minipage}
\end{table}
\noindent Experimental results show that LatentTune significantly outperforms existing baselines, achieving up to 1332\% improvement in RocksDB and 11.82\% throughput 46.01\% latency gains in MySQL. In future work, we plan to extend LatentTune to support real-time tuning in dynamic environments, moving closer to practical deployment in production database systems.

\section*{Acknowledgment}
This research was supported by the National Research Foundation (NRF) funded by the Korean government (MSIT) (No. RS-2023-00229822).

\bibliographystyle{IEEEtran}
\bibliography{ref}

\end{document}